\documentclass[aps,prl,twocolumn,preprintnumbers,groupedaddress]{revtex4-1}

\usepackage{graphicx}
\usepackage{dcolumn}
\usepackage{bm}

\usepackage{amsfonts,amsmath}
\pdfoutput=1

\usepackage{color}
\usepackage{tabularx}

\usepackage{ifthen}
\usepackage{enumerate}
\usepackage{amsmath}
\usepackage{amssymb}
\usepackage[mathscr]{eucal}
\usepackage[errorshow]{tracefnt}
\usepackage{amsmath}
\usepackage{slashed}
\usepackage{amssymb}
\usepackage{array}
\usepackage{amsthm}
\usepackage{eucal}
\usepackage{epsf}
\usepackage{epsfig}
\usepackage{psfrag}
\usepackage{multirow}
\usepackage{wasysym} 
\usepackage{tikz-cd} 
\usepgfmodule{shapes}
\usetikzlibrary{backgrounds, arrows,calc,shapes,decorations.pathreplacing, decorations.markings, automata,positioning}

\newcommand{\nn}{\nonumber}
\newcommand{\rf}{r_{\phi}}

\newcommand{\beqa}{\begin{eqnarray}}
\newcommand{\eeqa}{\end{eqnarray}}

\def\W{{\cal W}} 

\newcommand{\tr}{\text{tr}\,}

\newcommand{\beq}{\begin{equation}}
\newcommand{\eeq}{\end{equation}}
\newcommand{\bea}{\begin{eqnarray}}
\newcommand{\eea}{\end{eqnarray}}

\newcommand{\CB}{{\mathcal B}}

\newcommand{\CF}{{\mathcal F}}

\newcommand{\CM}{{\mathcal M}}
\newcommand{\M}{{\mathfrak M}}
\newcommand{\CN}{{\mathcal N}}
\newcommand{\CO}{{\mathcal O}}

\newcommand{\CT}{{\mathcal T}}

\newcommand{\CW}{{\mathcal W}}
\newcommand{\CZ}{{\mathcal Z}}

\newcommand\qt{\tilde q}

\newcommand\pt{\tilde p}
\newcommand\Qt{\tilde Q}
\newcommand\Pt{\tilde P}

\newcommand{\be}{\begin{equation}}
\newcommand{\ee}{\end{equation}}

\newcommand{\bpic}{\begin{tikzpicture}}
\newcommand{\epic}{\end{tikzpicture}}

\newcommand{\UNfour}{U(1)\!-\![2]_{\CN\!=4}}

\newcommand{\SUoneF}{SU(2)}

\def\+{{+\!\!\!+}}

\def\a{\alpha} 
\def\b{\beta}

\def\0{\nonumber}

\def\W{{\cal W}}

\begin{document}

\preprint{SISSA 27/2017/MATE-FISI}

\title{Supersymmetric gauge theories with decoupled operators and chiral ring stability}


\author{Sergio Benvenuti}
\email{benve79@gmail.com}
\affiliation{International School of Advanced Studies (SISSA) \& INFN, Sezione di Trieste} 

\author{Simone Giacomelli}
\email{sgiacome@ictp.it}
\affiliation{International Center for Theoretical Physics \& INFN, Sezione di Trieste}


\date{\today}

\begin{abstract}
We propose a general way to complete supersymmetric theories with operators below the unitarity bound, adding gauge-singlet fields which enforce the decoupling of such operators. This makes it possible to perform all usual computations, and to compactify on a circle. We concentrate on a duality between an $\CN=1$ $SU(2)$ gauge theory and the $\CN=2$ $A_3$ Argyres-Douglas \cite{Maruyoshi:2016tqk, Maruyoshi:2016aim}, mapping the moduli space and chiral ring of the completed $\CN=1$ theory to those of the $A_3$ model. We reduce the completed gauge theory to $3d$, finding a $3d$ duality with $\mathcal{N}=4$ SQED with two flavors. The naive dimensional reduction is instead $\mathcal{N}=2$ SQED. Crucial is a concept of chiral ring stability, which modifies the superpotential and allows for a $3d$ emergent global symmetry.
\end{abstract}


\maketitle


In gauge theories with four supercharges, many non-perturbative properties of the infrared strongly coupled fixed points are known, for instance the scaling dimensions of $\frac{1}{2}$-BPS operators, i.e. operators in the chiral ring \cite{Intriligator:2003jj, Kapustin:2009kz, Jafferis:2010un}. This provides invaluable insights about generic properties of renormalization group (RG) flows in quantum field theories. Any new non-perturbative method to analyze supersymmetric RG flows is of general interest.

 Sometimes a BPS operator violates the bound imposed by conformal invariance and unitarity, which in $4d$ ($3d$) is $\Delta >1$ ($\Delta >\frac{1}{2}$). The standard lore is that the operator decouples and becomes free \cite{Kutasov:2003iy}: the infrared fixed point is described by some interacting superconformal theory (SCFT) plus a free chiral field. How to perform computations in such theories is however an open problem: it is known how to perform $a/\CZ$-extremizations \cite{Intriligator:2003jj, Kapustin:2009kz,Jafferis:2010un} or compute supersymmetric indices/partition functions, but it is not known how to compute, for instance, the chiral ring or the moduli space of vacua.

In this note we propose a prescription to re-formulate theories with decoupled operators: introduce a gauge-singlet chiral multiplet $\b_{\CO}$ for each operator $\mathcal{O}$ violating the bound, and add the superpotential term $\b_{\small {\CO}}\mathcal{O}$. Gauge singlet fields entering the superpotential in this way are usually said to ``flip the operator $\mathcal{O}$''. The $\CF$-term of $\b_{\CO}$ sets $\CO=0$ in the chiral ring, there are no unitarity violations and all usual computations can be performed. 

This "completion" isolates the interacting sector and also allows to compactify dualities where at least one side has decoupled operators. Unitarity bounds change as we change the dimension of spacetime and what decouples in higher dimension may not decouple in lower dimension, so a compactification of dual theories without introducing the $\b_\CO$ fields generically fails to produce a dual pair. 

We check the validity of our proposal focusing on a class of  theories in four dimensions recently discovered in \cite{Maruyoshi:2016tqk, Maruyoshi:2016aim, Agarwal:2016pjo}: certain $\CN=1$ gauge theories exhibit unitarity bound violations, the interacting sector is proposed to be equivalent to a well-known class of $\CN=2$ SCFT's called Argyres-Douglas (AD) theories \cite{Argyres:1995jj,Argyres:1995xn,Eguchi:1996vu,Eguchi:1996ds}, which cannot have a manifestly $\CN=2$ lagrangian description.

We focus on a simple case, the $A_3$ AD theory, which admits an $\CN=1$ lagrangian description in terms of an $SU(2)$  gauge theory with an adjoint and two doublets \cite{Maruyoshi:2016aim}. 

First we point out that the superpotential as written in \cite{Maruyoshi:2016aim,Agarwal:2016pjo} are inconsistent: a superpotential term must be discarded, in order to satisfy a \emph{chiral ring stability} criterion as in \cite{Collins:2016icw}. Our consistent superpotential displays the correct global symmetry and allows to map the moduli space of vacua and the chiral ring across the duality.

The naive dimensional reduction of the $4d$ RG flow described in \cite{Maruyoshi:2016aim} lands on $\CN=2$ SQED with two flavors. The dimensional reduction of the version with the $\b_\CO$ fields instead flows to $\CN=4$ SQED with two flavors, displaying enhanced supersymmetry.

We generalize this example to $SU(N)$ (dual to $A_{2N-1}$ AD) in \cite{Benvenuti:2017kud}, where we also  discuss the $3d$ mirrors.

\section{Unitarity bounds and flipping fields}
\cite{Maruyoshi:2016aim} started from the $4d$ theory $\CN=2$ $SU(2)$ SQCD with $8$ doublets, applied a certain procedure that breaks half of the supersymmetry, and in the IR found an $\CN=1$ lagrangian: a $SU(2)$ gauge theory with an adjoint $\phi$, two doublets ($q, \qt$) various singlets and superpotential
\be\label{MSir}\W=tr(\tilde{q}\phi^4q) + \alpha_0tr(\tilde{q}q)+ \alpha_1tr(\tilde{q}\phi q)+ \alpha_2 tr(\tilde{q}\phi^2q)\ee 
A-maximization for theory (\ref{MSir}) shows that the operators $tr(\phi^2),\a_1,\a_2$ violate the unitarity bound $R\!>\!R_{\small bound}\!=\!\frac{2}{3}$. 

In theories where an operator $\CO$ violates the bound, it is customary to compute $a,c$ central central charges subtracting the contribution of a chiral field with $R$-charge $R[\CO]$. Similarly, partition functions are computed dividing by the contribution of a chiral field with $R$-charge $R[\CO]$. Using these recipes, the leftover interacting sector in (\ref{MSir}) was argued to be dual to $A_3$ AD theory, with $\alpha_0$ mapping to the Coulomb branch generator \cite{Maruyoshi:2016aim}. 

Let us emphasize that the above recipes do not give a complete description of the theory. Our proposal to deal with such cases is:

\emph{for each chiral ring operator $\CO$ with $R[\CO]\!\leq\!R_{\small bound}$, add to the theory a gauge invariant chiral multiplet $\b_{\CO}$, and add to the superpotential a "flipping" term $\delta\CW=\b_{\CO} \cdot \CO$.}

This formulation of the theory is complete, the $\CF$-terms of $\b_\CO$ imply that $\CO$ is zero in the chiral rings, so there are no unitarity violations, moreover all standard computations can be performed. 

As far as the computations which are possible just stating that $\CO$ is decoupled, our procedure is equivalent, as we now show.

The $4d$ $a$ and $c$ central charges are certain linear combinations of the cubic 't Hooft anomalies $tr(R)$ and $tr(R^3)$ \cite{Anselmi:1997am} (traces are over all elementary fermions).
Since \be \label{Rrel}R[{\b_\CO}]=2-R[\CO]\,,\ee
 adding to $tr(R^3)$ (or to $tr(R)$) a contribution of $\b_{\CO}$ is equivalent to subtracting the contribution of $\CO$. So $tr(R^3), tr(R), c$ and $a$ are the same. This is also true at the level of trial central charge, when performing a-maximization, so the prescription proposed in \cite{Kutasov:2003iy} to implement the decoupling gives the same result. 
 
 Similarly,  the $4d$ superconformal index obtained with our prescription is the same: denoting by $\Gamma_{ell}(r)$ the contribution of a chiral field with R-charge $r$, we have $\Gamma_{ell}(r) \Gamma_{ell}(2-r)=1$. So adding the contribution of $\b_{\CO}$, i.e. multiplying the index by $\Gamma_{ell}(R[\b_\CO])$, is equivalent to removing the contribution of $\CO$, i.e. dividing by $\Gamma_{ell}(R[\CO])$. Analogous arguments hold for $3d$ superconformal indices or $S^3$ partition functions.

\paragraph{Example} As a simple check of our proposal, let us consider in $4d$ $\CN=2$ $SU(N_c)$ Super Yang-Mills: it is well known that in the IR it is described by $N_c-1$ free $\CN=2$ vector multiplets. In $\CN=1$ terms, the theory is $SU(N_c)$ with an adjoint field $\phi$ and $\CW=0$. The NSVZ beta function is proportional to $N_c \rf$, where $\rf$ is the R-charge of the adjoint field. So an IR fixed point requires $\rf=0$: the $N_c-1$ operators $tr(\phi^j)$, $j=2,3,\ldots,N$ violate the unitarity bound and become free $\CN=1$ chiral multiplets. Adding the corresponding flipping fields $\b_j$
$$\CW= \sum_{j=2}^N \b_j tr(\phi^j)$$
%
we isolate the remaining sector. Since gauginos have R-charge $1$, the contribution to $tr(R^3)$ and $tr(R)$ of a chiral field of R-charge $2$ is the same of the contribution of an $\CN=1$ vector multiplet. Since $\rf=0$, in the $a$, $c$ central charges, the contribution of the gauginos cancel the contribution of the adjoint field $\phi$, the left-over is the $\b_j$'s, which, having R-charge $2$, are equivalent to $N_c-1$  free $\CN=1$ vector multiplets. So our prescription predicts the correct amount of free fields.

\section{Chiral ring stability}
Applying the above prescription to (\ref{MSir}), we need to add one flipping field, that we call $\b_2$, for $tr(\phi^2)$ ($\a_1,\a_2$ trivially decouple), so we get
\be\label{MSir1}\W=tr(\tilde{q}\phi^4q) + \alpha_0tr(\tilde{q}q) + \b_2 tr(\phi^2)\ee 
We claim that this superpotential is inconsistent, since it does not satisfy a criterion of \emph{chiral ring stability}.

The data that define a gauge theory are the matter content and the full Lagrangian. With four supercharges, usually the data are only the matter content and the superpotential, which is used to compute protected quantities like R-charges, chiral rings etc. If a superpotential does not satisfy chiral ring stability, problematic terms must be dropped (which means that they belong to the Kahler, unprotected, part of the Lagrangian), keeping such terms when computing protected quantities generally leads to wrong results, so superpotentials violating the chiral ring stability criterion are inconsistent.

First, let us state the criterion. Starting from a theory $\CT$ with superpotential $\CW_\CT = \sum_i \CW_i$ (where each term $\CW_i$ is gauge invariant), one needs, for each $i$, to:
\begin{itemize}
\item consider the modified theory $\CT_i$, where the term $\CW_i$ is removed from $\CW$
\item check if the operator $\CW_i$ is in the chiral ring of $\CT_i$
\end{itemize}
If one of the terms $\CW_i$ is not in the chiral ring of $\CT_i$, it must be discarded from the full superpotential $\CW_\CT$.

Notice that we are not requiring that $\CW_i$ is a relevant deformation of $\CT_i$, only that it is in the chiral ring of $\CT_i$.

A similar criterion of chiral ring stability was formulated in \cite{Collins:2016icw}, where it was also linked to the algebro-geometric criterion of K-stability. The difference w.r.t. \cite{Collins:2016icw} is that we are not considering the central charges of the "test chiral rings".  It would be interesting to study in more detail this procedure, and if our slightly simplified approach is equivalent to the one of \cite{Collins:2016icw}.

\paragraph{Dropping the superpotential term}
Let us study the superpotential (\ref{MSir1}). As any traceless $2 \times 2$ matrix, the adjoint field $\phi$ satisfies the algebraic relation $\phi^2=-det(\phi)\mathbb{I}_{2\times 2}$, which implies $tr(\qt \phi^4 q)= \frac{1}{4} tr(\phi^2) tr(\qt q)$. Let us now consider the modified theory obtained by removing the term $tr(\qt \phi^4 q)$ from (\ref{MSir1}). In the modified theory, $tr(\qt \phi^4 q)$ vanishes in the chiral ring, either for the $\CF$-terms of $\a_0$ ($tr(\qt q)=0$), or for the $\CF$-terms of $\b_2$ ($tr(\phi^2)=0$). So $tr(\qt \phi^4 q)$ does not pass the test of chiral ring stability and must be discarded. The correct superpotential is 
\be \label{stableIR}\CW_{\text{stable}}= \a_0 tr (\qt q) + \b_2 tr(\phi^2) \ee
It turns out that the relation imposed on the R-charges by $tr(\qt \phi^4 q)$ is equal to the relation from the NSVZ $\beta$-function, so a-maximization for theories (\ref{stableIR}) and (\ref{MSir1}) gives the same answer. $tr(\qt \phi^4 q)$ would however break an $SU(2)_F$ global symmetry necessary for the mapping to the $A_3$ AD model, moreover there is a crucial difference between (\ref{stableIR}) and (\ref{MSir1}) when compactifying to $3d$. Our two modifications of the Lagrangians of \cite{Maruyoshi:2016tqk, Maruyoshi:2016aim} make all the properties of the IR fixed point evident, in $4d$ and in $3d$.

\paragraph{Double dualities} We can provide a check of the validity of the chiral ring stability procedure using a "double duality": dualize a theory $\CT$ and get $\tilde{\CT}$, then dualize $\tilde{\CT}$ again, getting $\tilde{\tilde{\CT}}$. Usually $\tilde{\tilde{\CT}}=\CT$, but it might happen that some fields in the dual theory $\tilde{\CT}$ can be integrated out, and as a consequence $\tilde{\tilde{\CT}}$, compared to $\CT$, lacks some superpotential terms. The terms that disappeared in the double duality should be those which violate chiral ring stability. Let us show an example of this phenomenon in $3d$: start from $U(1)$ $\CN=2$ with $2$ flavors $p_1,p_2$ and 
\be\label{examp0} \W= \phi_1 p_1\pt_1 + \phi_2 p_2\pt_2 + \phi_2^{\,2} p_1\pt_1\,. \ee 
The last term does not satisfy the stability condition: if we remove $ \phi_2^{\,2} p_1\pt_1$, in the modified theory the $\CF$-terms of $\phi_1$ imply that $ \phi_2^{\,2}p_1\pt_1=0$ in the chiral ring.

Let us now "double dualize" (\ref{examp0}) as it is. The $3d$ mirror dual \cite{Aharony:1997bx} of $U(1)$ with two flavors and $\CW=0$ is $U(1)$ with two flavors $Q_1,Q_2$ and $\CW=\Phi_1 Q_1\Qt_1 + \Phi_2 Q_2\Qt_2$. $p_i\pt_i$ map to the singlets $\Phi_i$, so the $3d$ mirror dual of (\ref{examp0}) is $U(1)$ with two flavors and
\be\tilde{\CW}= \Phi_1 Q_1\Qt_1 + \Phi_2 Q_2\Qt_2 + \phi_1 \Phi_1 + \phi_2 \Phi_2 + \phi_2^{\,2} \Phi_1\,. \ee
Integrating out the massive fields leaves $\tilde{\CW}=0$. Taking the mirror again, we go back to 
$U(1)$ with $2$ flavors and $$\W = \phi_1 p_1\pt_1 + \phi_2 p_2\pt_2 \, :$$ the inconsistent term $ \phi_2^{\,2} p_1\pt_1$ in (\ref{examp0}) indeed disappears.


\section{Complete $4d$ chiral ring}
Let us go back to our $SU(2)$ gauge theory example
\be \CW_{\text{stable}}= \a_0 tr (\qt q) + \b_2 tr(\phi^2) \ee
Since we removed $tr(\qt \phi^4 q)$, there is a non abelian global symmetry $SU(2)_F$ under which $\{q,\qt\}$ is a doublet. The gauge invariant $tr(\qt q)$ is a $SU(2)_F$-singlet, while $\{\varepsilon_{a b} q^a\,  (\phi^{r} q)^b, tr(\qt \phi^r q), \varepsilon^{a b}  ( \qt \phi^{r})_a \, \qt_b\}$ form a $SU(2)_F$-triplet. There is also an abelian symmetry $U(1)_T$ that mixes with $U(1)_R$. A-maximization gives
\be\label{char21}
\begin{array}{c|ccc}
 &  U(1)_{R}^{4d} &  U(1)_T  & SU(2)_F  \\ \hline
\phi & \frac{2}{9} &  \frac{2}{9}  & {\bf 1}  \\
q,\qt & \frac{5}{9} &  -\frac{4}{9} &  {\bf 2} \\ 
\b_2 &  \frac{14}{9} & - \frac{4}{9} & {\bf 1} \\ 
\a_0 &  \frac{8}{9} & \frac{8}{9} &   {\bf 1} \\ 
\end{array}
\ee
where we normalized the $U(1)_T$ so that $R[\phi]=T[\phi]$. 

Let us try to give a vev to $\b_2$: $\phi$ becomes massive, and the IR theory is $SU(2)$ with two doublets and superpotential $\CW=\a_0 tr(\qt q)$. This theory quantum mechanically generates a ADS superpotential, and has no vacuum. We conclude that $\b_2$ cannot take a vev for quantum reasons. 

Since $\phi^2=0$ in the chiral ring, as a $2 \times 2$ matrix, the dressed mesons $tr(\qt \phi^{r_1} q)$ and dressed baryons $\varepsilon_{a b} \, (\phi^{r_2} q)^a\,  (\phi^{r_3} q)^b$ vanish in the chiral ring if any $r_i>1$.

The moduli space of vacua is described in terms of only four gauge invariant operators: $\a_0$ (with $\Delta=\frac{4}{3}$ and $R=T$) and the $SU(2)_F$-triplet $\{\CB=\varepsilon_{ab}\, q^a (\phi q)^b, \tilde{\CB}=\varepsilon^{ab} \, (\qt \phi)_a \qt_b, \CM=tr(\qt \phi q)\}$ (with $\Delta=2$ and $R=-2T$). 

The $SU(2)_F$-triplet satisfies the relation
\be \CB \tilde{\CB} = \varepsilon_{ab} \, q^a (\phi q)^b \varepsilon^{cd}  \, (\qt \phi)_c \qt_d = \CM^2 \ee
(we used that $tr(\qt q)=0$ in the chiral ring), which is the defining equation of $\mathbb{C}^2/\mathbb{Z}_2$, known to be the Higgs branch of $A_3$ AD. $\a_0$ maps to the Coulomb branch generator of $A_3$ AD \cite{Maruyoshi:2016aim}. The product between $\a_0$ and anyone among $\CB,\tilde{\CB},\CM$ vanishes, due the $\CF$-terms of $q$ and $\qt$.
We conclude that (\ref{stableIR}) and $A_3$ AD CFT have the same moduli space of vacua: a one-complex dimensional Coulomb branch and a two-complex dimensional Higgs branch intersecting at the origin.

We also propose that the holomorphic operator $\b_2$, which we saw cannot take a vev, is the superpartner of the operator $\a_0$ under the emergent $\CN=2$ supersymmetry. $\frac{1}{2}$-BPS $\CN=2$ multiplets for Coulomb branch operators indeed contain two complex scalars, with dimension $\Delta$ and $\Delta+1$, respectively. (For instance, in the $\CN=2$ SCFT $SU(2)$ with $4$ flavors, they are $tr(\Phi^2)$ with $\Delta=2$ and $tr(W_\a^2)$ with $\Delta=3$, the latter cannot take a vev.) From (\ref{char21}) we indeed see $R[\b_2]=R[\a_0]+\frac{2}{3}$. This fact also implies that $\b_2$ vanishes in the chiral ring.

\section{Down to $3d$: Abelianization}
Let us compactify the $4d$ story to $3d$. The $3d$ mirror \cite{Intriligator:1996ex} of $A_3$ AD reduced to $3d$ has been proposed \cite{Nanopoulos:2010bv, Boalch} to be the $\CN=4$ supersymmetric $U(1)$ with $2$ flavors $P_1,P_2$, 
\be \CW_{\UNfour}= \Phi (P_1\Pt_1+P_2\Pt_2) \ee
This theory, that we denote $\UNfour$, happens to be self-mirror, so we expect to find $\UNfour$ also analyzing the compactification on a circle of our theory (\ref{stableIR}), which is $3d$ $SU(2)$ with two doublets and one adjoint
 \be \label{Wstable1} \CW_{3d, \text{stable}}= \a_0 tr (\qt q) + \b_2 tr(\phi^2) \ee
Also in $3d$ $\b_2$ is not in the chiral ring: if we give a vev to $\b_2$, $\phi$ becomes massive, and the theory is $SU(2)$ with $1$ flavor. $SU(2)$ with $1$ flavor in $3d$ is described by a quantum modified moduli space \cite{Aharony:1997bx} $\M_{SU(2)}\, tr(\qt q)=1$ (where $\M_{SU(2)}$ is the basic monopole with GNO charges $\{+1,-1\}$), which is inconsistent with the $\CF$-terms of $\a_0$. 

No monopole superpotential is generated compactifying on $S^1$, since the only superpotential term that can soak up the zero modes of the adjoint field $\phi$ is $\b_2 tr(\phi^2)$, generating $\b_2\M_{SU(2)}$ \cite{Nii:2014jsa}. But a term $\b_2\M_{SU(2)}$ does not satisfy the chiral ring stability criterion. 

The absence of the terms $tr(\qt \phi^4 q)$ and $\b_2 \M_{SU(2)}$ is crucial: there is a $3d$ accidental $U(1)$ global symmetry 
\be\label{char3d}
\begin{array}{c|cccc}
 &  U(1)_{R} &  U(1)_q  & U(1)_{T'}  & SU(2)_{F}  \\ \hline
\phi & \rf & 0 & 1 & {\bf 1}  \\
q,\qt & r_q &  \frac{1}{2} & -\frac{1}{2} &  {\bf 2} \\ 
\b_2 &  2-2\rf & 0 & -2 &   {\bf 1} \\ 
\a_0 &  2-2r_q & -1 &  1 &  {\bf 1} \\ 
\end{array}
\ee
$T'$ is chosen so that the $SU(2)_F$-triplet is neutral. 

Let us study the $S^3$ partition function \cite{Kapustin:2009kz,Jafferis:2010un}. The contribution of a chiral multiplet with r-charge $r$ is $e^{l(1-r)}$, where $\partial_z l(z) =-\pi z cot(\pi z)$. The $S^3$ partition function for  $SU(2)$ with an adjoint of r-charge $r_{\phi}$ and a pair of doublets $q,\qt$ of r-charge $r_q$ is
\bea \CZ_{\SUoneF}[r_{\phi},r_q,b]= \frac{e^{l(-1+2r_q)}e^{l(-1+2r_{\phi})}}{2!}\cdot \qquad\qquad\\
\qquad \cdot \int_{-\infty}^{+\infty} \!\!\!\!\!\!e^{l(1-r_{\phi} \pm 2 i z)}e^{l(1-r_{\phi})}(2 sinh(2\pi z))^2e^{l(1-r_q \pm b \pm i z)} dz \nn \eea
$e^{l(-1+2r_q)}e^{l(-1+2r_{\phi})}$ is the contribution of the two singlets $\a_0$ and $\b_2$, and $b$ is the $SU(2)_F$ fugacity. 

We performed numerical $\CZ$-extremization in the two variables $\rf,r_q$. $\CZ_{\SUoneF}[r_{\phi},r_q,b]$ has a critical point at $r_q=\frac{1}{2} \, ,\rf=0$. (This result would have been in conflict with the superpotential terms $tr(\qt\phi^4q)$ or $\b_2\M_{SU(2)}$, which would impose the constraint $2=4\rf+2r_q$, and the numerical result would be $\rf=r_q=\frac{1}{3}$.)

Using the numerical input that $\rf=0$, we can see analytically that something interesting happens to the $S^3$ partition function. Since in the limit $\rf \rightarrow 0$,  $e^{l(1-r_{\phi} \pm 2 i z)}(2 sinh(2\pi z))^2= 1$ and $e^{l(-1 + 2 \rf) + l(1 - \rf))}=2$, the contributions from $\phi$ and $\b_2$, at $\rf\rightarrow 0$ cancel against the Haar measure and the Weil factor $2!$. The integrand in $\CZ_{\SUoneF}[r_{\phi}\rightarrow 0,r_q,b]$ reduces to the integrand of the partition function of $\UNfour$,
 \be \nn \CZ_{\UNfour}[r_P,B,\eta]\!=\!e^{l(-1+2r_P)}\!\!\!\int_{-\infty}^{+\infty} \!\!\!\!e^{l(1-r_P \pm B \pm i z)} e^{2 \pi \eta z} dz \,,\ee
computed at $\eta=0$. We interpret this reduction at the level of the integrands of the $S^3$ partition functions as a strong indication that \emph{ the full physical theory (\ref{Wstable1}) is equivalent to the Abelian theory $\UNfour$}.

If $\rf>0$, we checked numerically the equality among the $S^3$ partition functions in $3$ variables:
$$ \label{3variableq}\!\!\! \CZ_{\SUoneF}[\rf,r_q,b]\! =\! \CZ_{\UNfour}[r_P\!=\!r_q+\frac{\rf}{2},B\!=\!b,\eta\!=\!\rf] $$
The precise mapping between the $3$ variables can be understood from the mapping of the chiral rings, that we now proceed to study.

Compared to $4d$, in $3d$ there are additional chiral ring generators: the monopole operators, that can also be 'dressed' by the adjoint field.  Dressed monopoles have been studied for non-Abelian gauge theories in \cite{Cremonesi:2013lqa}, using Hilbert Series techniques \cite{Benvenuti:2006qr}. The analysis was done for $\CN=4$ gauge theories, but since the result is algebraic in nature, we can adapt it to our case. For $\CN=4$ $SU(2)$ gauge theories, the monopoles generating the Coulomb Branch are $\M_{SU(2)}$ and $\M_{SU(2)}$ dressed by one factor of $\phi$, that we denote $\{\M_{SU(2)}\phi\}$. In our $\CN=2$ case, these two monopoles pair up with $\a_0$ to form the $SU(2)_{\text{topological}}$ triplet generating the $3d$ Coulomb Branch of the Abelian theory:
\be\label{CBmap} \{\a_0,\{\M_{SU(2)}\phi\}, \M_{SU(2)} \} \leftrightarrow \{\M_{U(1)}^+,\Phi,\M_{U(1)}^-\}\ee
The $SU(2)_F$-triplet instead map to the Higgs Branch:
\be \{\CB,\CM,\tilde{\CB}\}  \leftrightarrow \{P_1\Pt_2, P_1\Pt_1\!-\!P_2\Pt_2, P_2\Pt_1\}\ee
Concluding, we showed that all the chiral generators map according to the Abelianization duality. We are unable to determine all the chiral ring relations, we notice that the duality predicts the relation $\a_0 \cdot \M_{SU(2)}=\{\M_{SU(2)}\phi\}^2$.

The UV global symmetries 
$( U(1)_R \times U(1)_q ) \times U(1)_{T'} \times SU(2)_{F}$
of the $SU(2)$ gauge theory enhance in the IR to 
$SO(4)_R \times SU(2)_{\text{topological}} \times  SU(2)_{\text{flavor}}$.

\subsection{Maruyoshi-Song procedure in $3d$: only $\CN=2$ susy}\label{MSin3d}
If we repeat the procedure of Maruyoshi and Song in $3d$, or if we naively reduce (\ref{MSir}) from $4d$, we find a theory with the same superpotential as in (\ref{MSir}):
\be \label{Wtrial}\CW_{\text{trial}}= tr(\qt \phi^4 q)+ \a_0 tr(\qt q)+ \a_1 tr(\qt \phi q)+ \a_2 tr(\qt \phi^2 q) \ee
First we need to consider chiral ring stability. Using the relation $\phi^2= -det(\phi) \mathbb{I}$ and the $\CF$-terms of $\a_0$ ($tr(\qt q)=0$), it is easy to see that both the first and last term in (\ref{Wtrial}) vanish in the modified chiral rings, so these two terms must be discarded.
$\CZ$-extremization in two variables now gives $\rf\simeq0.2088,\, r_q \simeq 0.4698$, so now $tr(\phi^2)$ violates the unitarity bound $r>\frac{1}{2}$: a singlet field $\b_2$ flipping $\tr(\phi^2)$ must be introduced. For the complete and stable theory 
\be\label{Wstable2} \CW = \a_0 tr(\qt q)+ \a_1 tr(\qt \phi q)+ \b_2 tr(\phi^2)\,,\ee
and $\CZ$-extremization gives $\rf=0, r_q \simeq 0.5918$. 

(\ref{Wstable2}) is dual to $\CN=2$ $U(1)$ with $2$ flavors $Q_1, Q_2$ and 
\be \label{UFF}\CW=\Phi_1 Q_1\Qt_1+\Phi_2 Q_2\Qt_2 \ee
The chiral ring generators of (\ref{Wstable2}) and (\ref{UFF}) map as follows 
$$\,\,\,\,\,\{ \CB, \tilde{\CB} \} \,\, \leftrightarrow \,\, \{Q_1 \Qt_2,Q_2\Qt_1\}$$
\vspace{-0.8cm}
$$\{\M,\{\M\phi\},\a_0,\a_1\} \leftrightarrow \{\M_{U(1)}^+,\Phi_1,\Phi_2,\M_{U(1)}^-\}\,\,\,$$
In summary, in order to compactify the $4d$ duality between $\CN=1$ gauge theory and the $\CN=2$ AD model, we first need to introduce the $\b_\CO$ flipping fields, then compactify. If we naively compactify the non-Abelian $\CN=1$ theory without the $\b_\CO$ fields, there is no enhancement of supersymmetry, i.e. the $4d$ dual pair does not descend to a $3d$ dual pair. This obstruction to compactification of dualities is unrelated to the emergence of monopole superpotentials discussed in detail in \cite{Aharony:2013dha}.
\vspace{1cm}

\begin{acknowledgments}

We are indebted to Amihay Hanany, Sara Pasquetti, Jaewon Song and especially Alberto Zaffaroni for discussions and helpful comments. S.B. is partly supported by the INFN Research Projects GAST and ST\&FI and by PRIN "Geometria delle variet\`a algebriche''. The research of S.G. is partly supported by the INFN Research Project ST\&FI.

\end{acknowledgments}

\bibliographystyle{ytphys}

\begin{thebibliography}{100}
  
\bibitem{Maruyoshi:2016tqk} 
  K.~Maruyoshi and J.~Song,
  ``The Full Superconformal Index of the Argyres-Douglas Theory,''
  arXiv:1606.05632 [hep-th].
  
\bibitem{Maruyoshi:2016aim} 
  K.~Maruyoshi and J.~Song,
  ``$ \mathcal{N}=1 $ deformations and RG flows of $ \mathcal{N}=2 $ SCFTs,''
  JHEP {\bf 1702}, 075 (2017)
  doi:10.1007/JHEP02(2017)075
  [arXiv:1607.04281 [hep-th]].

\bibitem{Intriligator:2003jj} 
  K.~A.~Intriligator and B.~Wecht,
  ``The Exact superconformal R symmetry maximizes a,''
  Nucl.\ Phys.\ B {\bf 667}, 183 (2003)
  doi:10.1016/S0550-3213(03)00459-0
  [hep-th/0304128].

\bibitem{Kapustin:2009kz} 
  A.~Kapustin, B.~Willett and I.~Yaakov,
  ``Exact Results for Wilson Loops in Superconformal Chern-Simons Theories with Matter,''
  JHEP {\bf 1003}, 089 (2010)
  doi:10.1007/JHEP03(2010)089
  [arXiv:0909.4559 [hep-th]].
  
\bibitem{Jafferis:2010un} 
  D.~L.~Jafferis,
  ``The Exact Superconformal R-Symmetry Extremizes Z,''
  JHEP {\bf 1205}, 159 (2012)
  doi:10.1007/JHEP05(2012)159
  [arXiv:1012.3210 [hep-th]].
     
\bibitem{Kutasov:2003iy} 
  D.~Kutasov, A.~Parnachev and D.~A.~Sahakyan,
  ``Central charges and U(1)(R) symmetries in N=1 superYang-Mills,''
  JHEP {\bf 0311}, 013 (2003)
  doi:10.1088/1126-6708/2003/11/013
  [hep-th/0308071].

\bibitem{Agarwal:2016pjo} 
  P.~Agarwal, K.~Maruyoshi and J.~Song,
  ``$ \mathcal{N} $ =1 Deformations and RG flows of $ \mathcal{N} $ =2 SCFTs, part II: non-principal deformations,''
  JHEP {\bf 1612}, 103 (2016)
  doi:10.1007/JHEP12(2016)103
  [arXiv:1610.05311 [hep-th]].

\bibitem{Argyres:1995jj}
  P.~C.~Argyres and M.~R.~Douglas,
  ``New phenomena in SU(3) supersymmetric gauge theory,''
  Nucl.\ Phys.\ B {\bf 448} (1995) 93
  doi:10.1016/0550-3213(95)00281-V
  [hep-th/9505062].

\bibitem{Argyres:1995xn}
  P.~C.~Argyres, M.~R.~Plesser, N.~Seiberg and E.~Witten,
  ``New N=2 superconformal field theories in four-dimensions,''
  Nucl.\ Phys.\ B {\bf 461} (1996) 71
  doi:10.1016/0550-3213(95)00671-0
  [hep-th/9511154].

\bibitem{Eguchi:1996vu}
  T.~Eguchi, K.~Hori, K.~Ito and S.~K.~Yang,
  ``Study of N=2 superconformal field theories in four-dimensions,''
  Nucl.\ Phys.\ B {\bf 471} (1996) 430
  doi:10.1016/0550-3213(96)00188-5
  [hep-th/9603002].

\bibitem{Eguchi:1996ds}
  T.~Eguchi and K.~Hori,
  ``N=2 superconformal field theories in four-dimensions and A-D-E classification,''
  In *Saclay 1996, The mathematical beauty of physics* 67-82
  [hep-th/9607125].

\bibitem{Collins:2016icw} 
  T.~C.~Collins, D.~Xie and S.~T.~Yau,
  ``K stability and stability of chiral ring,''
  arXiv:1606.09260 [hep-th].
  
\bibitem{Benvenuti:2017kud}
  S.~Benvenuti and S.~Giacomelli,
  ``Abelianization and Sequential Confinement in $2+1$ dimensions,''
  arXiv:1706.04949 [hep-th].

\bibitem{Anselmi:1997am} 
  D.~Anselmi, D.~Z.~Freedman, M.~T.~Grisaru and A.~A.~Johansen,
  ``Nonperturbative formulas for central functions of supersymmetric gauge theories,''
  Nucl.\ Phys.\ B {\bf 526}, 543 (1998)
  doi:10.1016/S0550-3213(98)00278-8
  [hep-th/9708042].

\bibitem{Aharony:1997bx} 
  O.~Aharony, A.~Hanany, K.~A.~Intriligator, N.~Seiberg and M.~J.~Strassler,
  ``Aspects of N=2 supersymmetric gauge theories in three-dimensions,''
  Nucl.\ Phys.\ B {\bf 499}, 67 (1997)
  doi:10.1016/S0550-3213(97)00323-4
  [hep-th/9703110].
 

\bibitem{Intriligator:1996ex} 
  K.~A.~Intriligator and N.~Seiberg,
  ``Mirror symmetry in three-dimensional gauge theories,''
  Phys.\ Lett.\ B {\bf 387}, 513 (1996)
  doi:10.1016/0370-2693(96)01088-X
  [hep-th/9607207].
  
\bibitem{Nanopoulos:2010bv}
  D.~Nanopoulos and D.~Xie,
  ``More Three Dimensional Mirror Pairs,''
  JHEP {\bf 1105} (2011) 071
  doi:10.1007/JHEP05(2011)071
  [arXiv:1011.1911 [hep-th]].

\bibitem{Boalch} 
P.~Boalch, 
``Irregular connections and Kac-Moody root systems,'' 
[arXiv:0806.1050 [math.DG]].
  
  \bibitem{Nii:2014jsa}
  K.~Nii,
  ``3d duality with adjoint matter from 4d duality,''
  JHEP {\bf 1502} (2015) 024
  doi:10.1007/JHEP02(2015)024
  [arXiv:1409.3230 [hep-th]].

\bibitem{Cremonesi:2013lqa} 
  S.~Cremonesi, A.~Hanany and A.~Zaffaroni,
  ``Monopole operators and Hilbert series of Coulomb branches of $3d$  $\mathcal{N} = 4$ gauge theories,''
  JHEP {\bf 1401}, 005 (2014)
  doi:10.1007/JHEP01(2014)005
  [arXiv:1309.2657 [hep-th]].
  
\bibitem{Benvenuti:2006qr} 
  S.~Benvenuti, B.~Feng, A.~Hanany and Y.~H.~He,
 ``Counting BPS Operators in Gauge Theories: Quivers, Syzygies and Plethystics,''
  JHEP {\bf 0711}, 050 (2007)
  doi:10.1088/1126-6708/2007/11/050
  [hep-th/0608050].


\bibitem{Aharony:2013dha}
  O.~Aharony, S.~S.~Razamat, N.~Seiberg and B.~Willett,
  ``3d dualities from 4d dualities,''
  JHEP {\bf 1307} (2013) 149
  doi:10.1007/JHEP07(2013)149
  [arXiv:1305.3924 [hep-th]].
  
  
  \end{thebibliography}

\end{document}